# A statistical analysis of pores and micro-cracks in nuclear graphite

Qing Huang, Hui Tang

Shanghai Institute of Applied Physics, Chinese Academy of Sciences, Shanghai 201800, China

Corresponding author: Qing Huang. Tel: 0086-021-39194239. E-mail: huangqing2012@sinap.ac.cn

**Abstract**

Microstructure characterization is of great value to understanding nuclear graphite's properties and irradiation behavior. However, graphite is soft and could be easily damaged during sample preparation. A three-step polishing method involving mechanical polishing, ion milling and rapid oxidation is proposed for graphite. Ion milling is adopted to remove the broken graphite pieces produced by mechanical polishing. Rapid oxidation is then adopted to remove irradiation-induced damage layer during ion milling. The Raman spectra show very low G peak width and $I_D/I_G$ ratio after rapid oxidation, indicating a surface completely free from artificial defects. The micro-cracks which were conventionally observed via a transmission electron microscope can be observed on rapid-oxidized surface in a scanning electron microscope. By digital image processing, the micro-cracks along with the gas-escape pores in nuclear graphite IG-110 are statistically analyzed. Porosity's distributions on crack (pore) size (spanning from 10 nm to 100 μm) are given, which could help to understand and simulate graphite's performances in reactors.

# 1. Introduction

Graphite materials have a variety of applications in many fields of industry and engineering. Particularly, highly purified graphite is proposed to be the core structural material in two types of Generation IV reactors: molten salt reactors and very high temperature reactors. The artificial graphite is usually manufactured from a filler coke and pitch binder. The release of volatile content of binder region during carbonization leaves gas-escape pores in the graphite product. Cooling down from high-temperature graphitization creates micro-cracks between graphite basal planes.

Both gas-escape pores and micro-cracks have a great influence on nuclear graphite's properties and performances. Graphite porosity is one of the most important factors that determine its oxidation behavior. Oxidation then changes graphite pore structure in return. In molten salt reactors, graphite components are in direct contact with molten salt. The salt fuel could intrude into graphite pores, creating hot spots in graphite. In order to prevent salt intrusion, the gas-escape pores should have sizes as small as ~1 μm. The micro-cracks between graphite sheets provide spaces for c-axis thermal expansion during heating, resulting in a good thermal shock resistance of graphite. Besides, during irradiation in reactors, the micro-cracks can accommodate the c-axis swelling of graphite crystallites, slowing down the accumulation of inner stress and macroscopic volume expansion.

A thorough characterization of microstructures and porosity of one graphite grade is very important for understanding its properties and performances in reactors. Furthermore, given the need of simulating and predicting nuclear graphite's behavior, the shapes and size distributions of graphite pores and micro-cracks should be known as input data for simulation. However, graphite microstructures are very complicated. The gas-escape pores and micro-cracks show various shapes, orientations and sizes in nuclear graphite. Direct observation and statistical analysis of graphite gas-escape pores have been achieved for pristine graphite [1-3], oxidized graphite [4,5] and irradiated graphite [6]. The X-ray tomography is a powerful technique to make a nondestructive analysis of large graphite pores (with the best resolution of around 1 μm) [3,5,6]. Digital image processing combined with optical microscopy has also been used to analyze the pore structures of various graphite grades [1,2,4]. The X-ray tomography and optical micrograph processing has been demonstrated to give consistent porosity distributions for IG-110 graphite [2]. A polishing process, usually mechanical polishing, is needed to observe graphite pores in a microscope. It was reported that mechanical polishing could greatly enhance the intensity of Raman D mode, indicating a remarkable increase of defects in graphite [7]. Therefore, the submicron-sized structures (such as micro-cracks) cannot be seen on mechanically-polished graphite surface. The micro-cracks conventionally were observed via transmission electron microscope (TEM) with a very limited field of view. The TEM specimens were prepared by low energy argon ion beam milling or focused gallium ion beam milling. Ion-milling-induced damage in alloys have been studied and successfully removed by an electropolishing method [8]. As for graphite, ion milling's influence to the microstructures has not been identified. A low-density carbon material has been found and studied during TEM observation of nuclear graphite specimens prepared by argon ion milling [9,10] or focused gallium ion milling [11,12]. But the origin of this carbon material has not been well explained. In recent years, the focused ion beam-scanning electron microscopy (FIB-SEM) tomography [12,13] and electron tomography [14] has been used to reconstruct the 3D structure of graphite micro-cracks. The FIB-SEM tomography demonstrated a significant development of pores with volume <0.1 $\mu m^3$ after high dose irradiation [12]. However, the sizes of samples for FIB-SEM tomography and electron tomography are limited to several microns. A statistical analysis of graphite micro-cracks is still a challenge.

The above-mentioned sample processing methods (mechanical polishing and ion milling) are able to damage the very surface of graphite specimens and bring some uncertainties to the analysis of graphite microstructures. In this study, the damages induced by mechanical polishing and ion milling are identified and effectively removed to produce a graphite surface without any artificial defect, so that the graphite micro-cracks on the sample's surface can be observed in a scanning electron microscope (SEM). A statistical analysis of graphite micro-cracks is then conducted.

## 2. Experimental

One typical nuclear grade of graphite, IG-110, is used in this study. The IG-110 graphite, from Toyo Tanso Inc., is produced from a petroleum coke with an average grain size of 20 μm. A sample with sizes of 15×15×15 mm was used in this study. The mass and dimensions of the samples were measured using a balance and micrometer respectively, in order to calculate the apparent density and total porosity.

A three-step polishing method is proposed in this study. The first step is mechanical polishing. In order to prevent the pores being blocked by graphite dust and polishing medium, a thermoplastic resin was used to fill the graphite pores before mechanical polishing. One facet of the sample was ground on a grit 600 silicon carbide abrasive paper and then was polished successively on cloth using 3 μm diamond and 20 nm silica suspensions for several minutes. After that, the sample was washed in acetone to remove the resin.

The second step is ion beam milling. A small sample with sizes of 3×3×2 mm was cut from the mechanically polished sample. The mechanically polished surface of the small sample was then bombarded by glancing-incident (6°) argon ion beams with energies of 4.5 and 3 keV successively in a Gatan Model 691 system. The sample size is limited by this ion milling system.

The third step is rapid oxidation. A furnace was electrically heated in air to 680 °C. Then open the furnace lid and put the ion-milled sample in the furnace. The ion-milled graphite surface was oxidized in air for three minutes. The sample was then taken out and cooled down in air. (It should be noted that the temperature was not specifically selected in this study. The furnace temperature was set to 750 °C. Due to its bad temperature probe, the actual temperature was calibrated by a thermal couple and was 680 °C. Besides, the temperature was not strictly controlled during oxidation since opening the furnace lid disturbed the temperature inside the furnace. The above mentioned operation of rapid oxidation is not rigid but very convenient and is good enough for the purpose of oxidation in this study.)

In order to analyze the gas-escape pores, the mechanically-polished surface of the sample was observed using an optical microscope with a magnification of 100×. The pixel resolution is 0.54 μm. In total 63 micrographs were recorded for the sample and were stitched together to form a large greyscale image (8×8 mm) of the polished surface. The pores in the optical images are shown as dark regions. Then, the pores were extracted from the optical images and were statistically analyzed.

The graphite surface was observed in a SEM after performing each polishing step. The element on the surface was measured by using the energy dispersive X-ray (EDX) spectroscopy. Especially, after rapid oxidation, the surface was observed in SEM with a magnification of 5000×. The pixel resolution is 7.5 nm. In total 48 micrographs were recorded for the sample and were stitched together to form a large greyscale image (0.13×0.13mm). Similar to the pores in optical image, the micro-cracks are shown as dark regions in the SEM image. The micro-cracks then were extracted from the image and were statistically analyzed.

In order to investigate the low-density carbon material reported in graphite micro-cracks, a thin disk specimen of IG-110 graphite was also prepared for TEM observation. A hole in the middle of the specimen was achieved by low energy argon ion milling which used the same ion energies as the second polishing step given above. The very thin area of the TEM specimen was also observed via SEM.

To study the damage induced by each polishing step, the Raman spectra were measured on sample's surface after mechanical polishing, ion beam milling and rapid oxidation. A Bruker SENTERRA confocal Raman microscope was used. The excitation light is a He-Ne laser with a wavelength of 633 nm. Five spots were randomly selected on the samples' surface for Raman spectra measurements.

## 3. Results and discussion

### 3.1 Identifying damage induced by mechanical polishing and ion milling

Fig. 1(a) shows a SEM image of mechanically-polished surface of IG-110 graphite. The widespread micro-cracks cannot be seen, and the fillers cannot be readily distinguished from the binder regions on the surface. One typical area is magnified in the inset image and shows small broken graphite pieces induced by friction between graphite surface and diamond particles during mechanical polishing.

Fig. 1(b) shows a SEM image of ion-milled IG-110 graphite. The broken graphite pieces have been completely removed. More importantly, micro-cracks (indicated by the white arrows in Fig. 1(b)) are observed on the ion-milled surface. However, it is observed that the very thin cracks are not empty after ion beam milling.

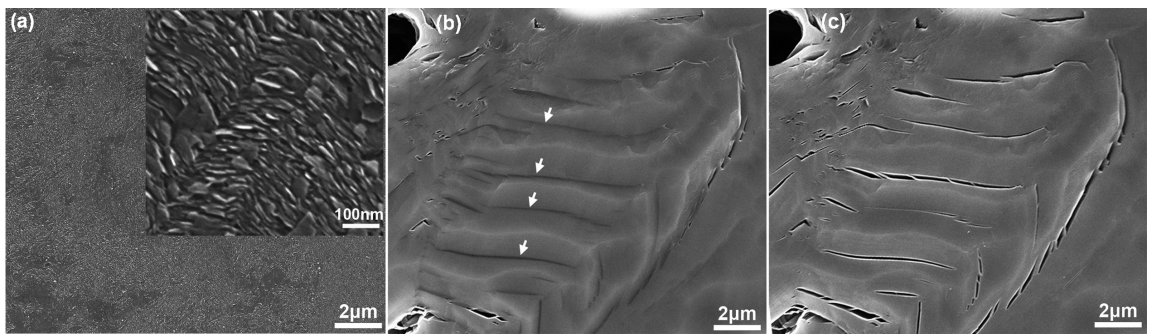

Fig. 1 (a) A SEM image of mechanically-polished surface of IG-110. One typical area is magnified in the inset image. (b) A SEM image of ion-milled surface of IG-110. The micro-cracks are indicated by white arrows (c) A SEM image of polished surface after rapid oxidation for 3 minutes.

In order to understand the origin of non-empty cracks, a thin disk specimen with a hole in the middle was prepared by low energy argon ion milling. Fig. 2(a) shows a TEM image of one filler in IG-110 graphite. The low-density carbon material which has been reported in previous studies [9-12] also appears in cracks shown in Fig. 2(a). If this carbon material is inherent in as-manufactured graphite, it should be found inside the cracks. The TEM image do show that the carbon material seems to reside inside the cracks. However, the cracks look different when observing them via SEM. The SEM image (Fig. 2(b)) shows that the cracks are highly likely to be covered by a thin layer of carbon material. The narrow cracks are fully covered and the wide cracks are partly covered. Especially the wide crack on the left is observed to be empty and partly covered by two thin layers of carbon material on the upper surface and the lower surface (indicated by arrows) respectively. Both sides of the thin specimen were processed by argon ion milling. Except for sputtering, another important effect is irradiation during argon ion milling. It is well known that irradiation induces swelling of graphite along the $c$ axis. The irradiation-induced $c$-axis swelling could be as high as 200% at room temperature [15]. Therefore, this carbon material should be caused by irradiation-induced $c$-axis swelling of graphite crystallites. The argon ion irradiation process was simulated by using "The Stopping and Range of Ions in Matter" program. The irradiation damage induced by 3-keV argon ions at a glancing angle of 6° is deposited in a 2-nm-thick surface layer. Thus the cracks shown in Fig. 2 should be covered by a 2-nm-thick layer of carbon material.

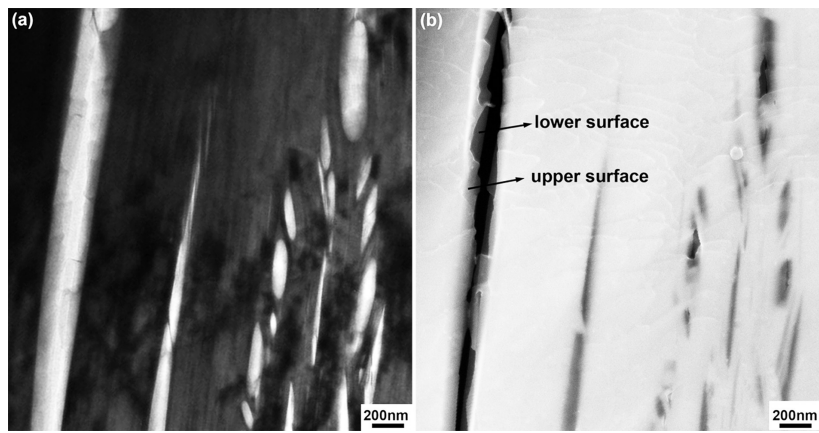

Fig. 2 (a) A TEM image of one filler with micro-cracks in IG-110 graphite and (b) a SEM image of the same filler.

The artificial carbon material induced by irradiation damage is apparently much more defective than the underlying graphite structure and is believed to be more susceptible to oxidation. Therefore a rapid oxidation process was adopted to remove this artificial carbon material. Fig. 1(c) shows a SEM image of IG-110 graphite after oxidation at 680 °C for 3 minutes. The artificial carbon material induced by argon ion milling is completely removed by rapid oxidation, and the micro-cracks can be clearly seen on the sample's surface.

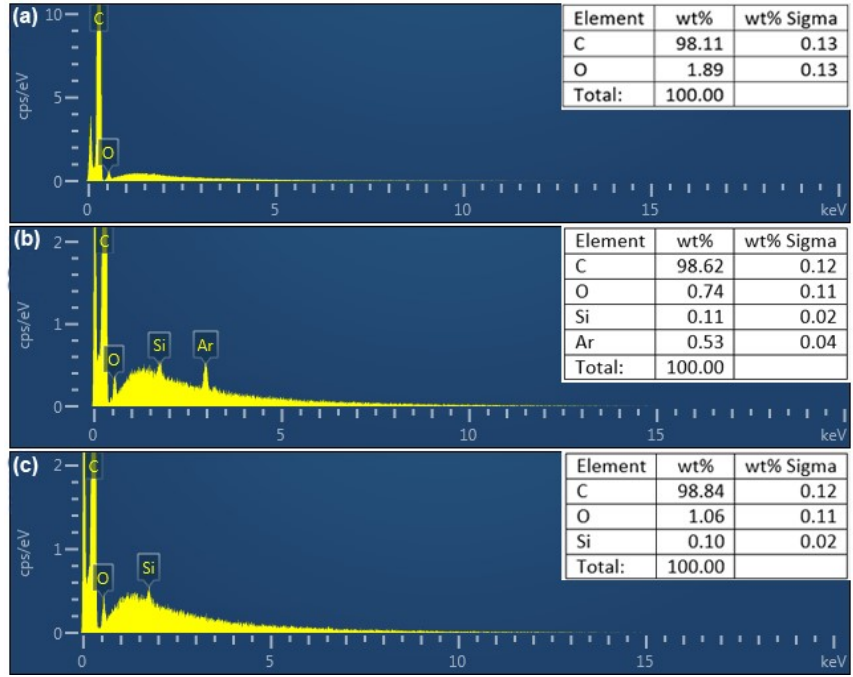

Fig. 3 The EDX spectra of (a) as-machined graphite surface and surface after (b) ion milling and (c) rapid oxidation.

Fig. 3(a) - 3(c) shows the EDX spectra of as-machined graphite surface and graphite surface after ion milling and rapid oxidation respectively. Oxygen, either chemically absorbed (C-O or C=O) or physically absorbed ($C(O_2)$), always can be detected on graphite surface even after ion milling. The reason is that ion milling only sputters the graphite surface and the graphite pores' inner surfaces were left untouched. After ion milling, a strong signal of argon can be seen in Fig. 3(b), indicating that the argon ions were implanted into graphite surface. Argon implantation has been found in highly ordered pyrolytic graphite surface sputtered with 5 keV argon ions [16]. As the argon ions slowed down and settled in graphite, argon-carbon collisions happened and induced irradiation damage, which supports above assumption that the artificial carbon thin layer was induced by irradiation damage. After rapid oxidation, the artificial carbon thin layer was removed and the argon was released. No argon signal was detected in Fig. 3(c).

It is worth mentioning that silicon has been found after ion milling and rapid oxidation.

The reason is that the sample was mechanically polished with silica suspensions before ion milling. We did not use ultrasonic cleaning to remove the silica particles adhering to the surface in order to prevent ultrasound-induced damage to graphite structures. The silica particles on the surface were then sputtered and redeposited on other surfaces, including the pores' inner surfaces. Silica particles in graphite pores were then detected during EDX spectrum measurements.

Fig. 4(a) shows the typical Raman spectra measured on IG-110 graphite surface after mechanical polishing, ion milling and rapid oxidation. Each spectrum has three peaks (the D peak located at ~1340 cm$^{-1}$, G peak at 1583 cm$^{-1}$ and the D' peak at 1619 cm$^{-1}$). The intrinsic G peak originated from the in-plane symmetric stretching motion of C-C sp$^2$ bond and is the only peak that can be detected from defect-free graphite crystal [17]. The D peak involves a breathing mode of six-fold rings of carbon atoms and only can be excited in the presence of in-plane defects [18].

Fitting of the Raman spectra was achieved by using three Lorentzian lines for the G, D and D' peaks. The peaks in the spectrum of ion-milled graphite seems to be raised up by a "bump" of the spectrum. Based on above discussion, ion beam milling introduced irradiation damage at surface and produced a highly damaged surface layer with a thickness of ~2 nm. The Raman spectrum of ion-milled graphite is composed of signals from the highly damaged thin layer (responsible for the "bump" of the spectrum) and the graphite structure underneath this layer. Therefore, one Gaussian line was added to fit the "bump" of the spectrum.

The full width at half maximum (FWHM) of the G peak and the ratio of the intensities of D and G peaks ($I_D/I_G$) has been widely used to analyze the microstructures of graphitic materials after various treatments [7,19]. The average value and the standard deviation (the error bar) of the FWHM and $I_D/I_G$ ratio are shown in Figs. 4(b) and 4(c) respectively.

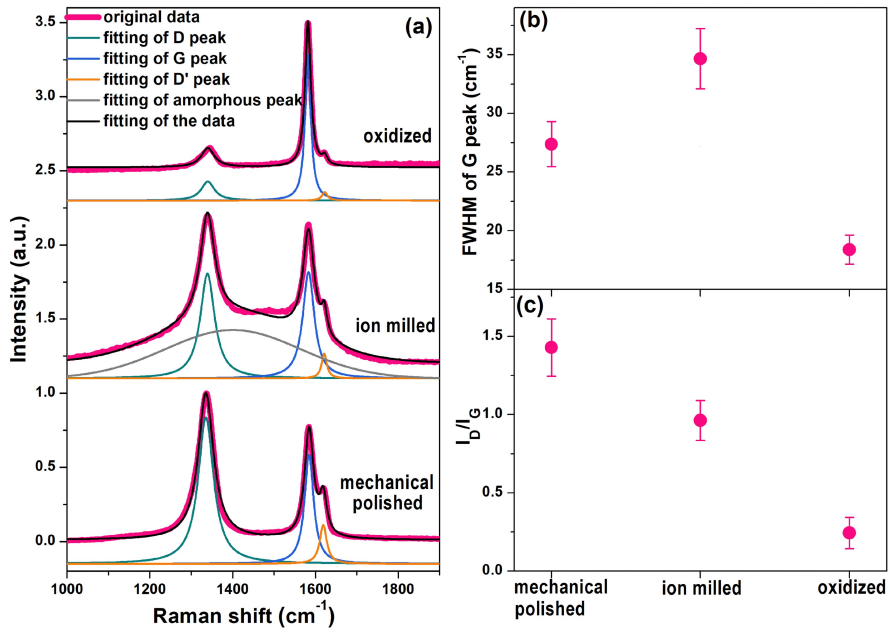

Fig. 4 (a) The Raman spectra and fitting curves of mechanically-polished, ion-milled and rapid-oxidized IG-110 graphite. (b) The FWHM of G peak. (c) The $I_D/I_G$ ratio.

The Raman spectra clearly show the damage induced by each polishing step. Mechanical polishing induced broken graphite sheets with abundant boundaries which could raise the D peak intensity. Fig. 4(c) shows that the mechanically-polished graphite surface shows the highest $I_D/I_G$ ratio (~1.4). Ion beam milling induced crystal disordering which causes a strong coupling between discrete G mode and a continuum of phonon states, resulting in peak broadening [20]. Therefore, the ion-milled graphite surfaces show the highest G peak width (Fig. 4(b)). After rapid oxidation, the G peak width and $I_D/I_G$ ratio reduced to 19 cm$^{-1}$ and 0.28 respectively.

It is worth mentioning that the $I_D/I_G$ ratio was reported to be inversely proportional to the in-plane graphite crystal size ($L_a$) [17,21], and this relationship has been used to obtain $L_a$ from the Raman spectra measured on graphite sample's surface. However, Fig. 4 shows that the mechanical machining, grinding or polishing could raise the $I_D/I_G$ ratio and thus underestimate the real $L_a$ of graphite crystallites inside the sample. The Raman spectra measured on rapid-oxidized surface are believed to be more appropriate for calculating the $L_a$ value.

### 3.2 Analysis of IG-110 graphite pores and micro-cracks

Fig. 5(a) is a SEM image of rapid-oxidized surface showing a typical filler particle in IG-110 graphite. There are abundant micro-cracks in this filler region. Some cracks are bridged by small graphite sheets (indicated by the two arrows). The graphite sheets are curved, deflected and folded in the filler. It is found that IG-110 is very rich in micro-cracks, which is consistent with its low coefficient of thermal expansion (4 × 10$^{-6}$ K$^{-1}$) and deep volume shrinkage (-5% at 750 °C) during neutron irradiation [22,23]. The quinoline insoluble (QI) particles can also be clearly seen on graphite's surface after rapid oxidation. Fig. 5(b) shows several QI particles observed at a magnification of 40000×. The nano-sized graphite sheets with their c axis directing radially toward the center of the QI particles are clearly seen. There was an attempt to identify the QI particles on mechanically-polished graphite surface [24]. It was observed that the typical structure of QI particles was damaged by mechanical polishing. The IG-110 graphite is characterized to have a limited number of QI particles. The agglomeration of QI particles which was reported in another nuclear graphite (NBG-18) [25] is rare to see in IG-110.

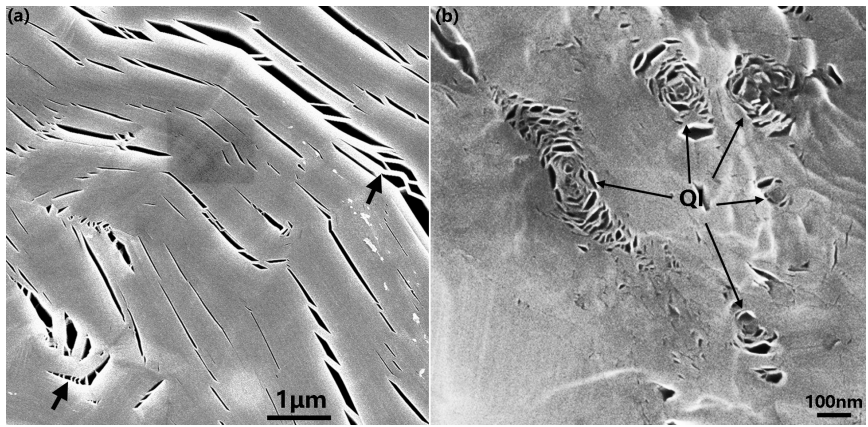

Fig. 5 The SEM images of ion-milled and rapid-oxidized IG-110 graphite surface. (a) A typical filler region showing abundant micro-cracks observed at a magnification of 5000×, and (b) the QI particles observed at a magnification of 40000×. The two arrows shown in (a) indicate graphite sheets bridging the micro-cracks.

In order to make a statistical analysis of gas-escape pores and micro-cracks, image stitching was used to achieve a micrograph with a wide field of view and a high resolution. A 8mm × 8mm optical micrograph (pixel resolution of 0.54 μm) of mechanically-polished graphite surface was obtained for pore structure analysis. A 0.13mm × 0.13mm SEM micrograph (pixel resolution of 7.5nm) of rapid-oxidized graphite surface was obtained for micro-crack analysis.

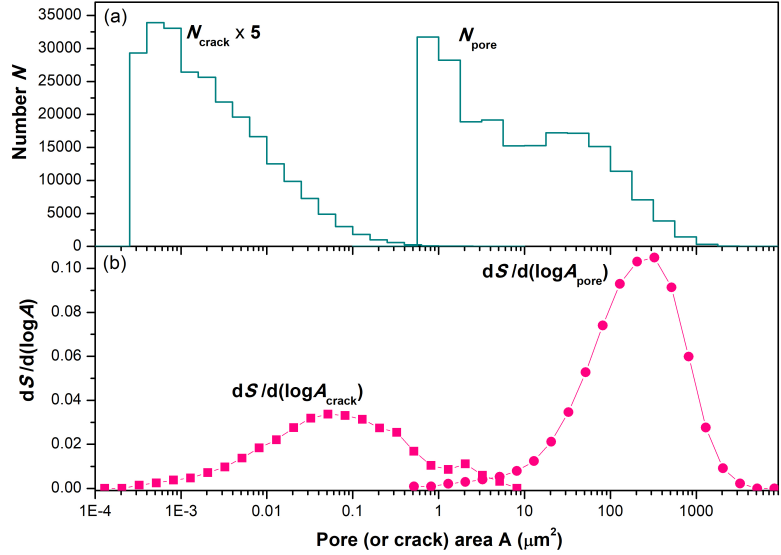

Fig. 6 (a) Distribution of pore (crack) area. (b) The porosity's distribution on pore (crack) area.

The pores and micro-cracks are shown as dark regions in the micrographs. A greyscale threshold value was selected to extract the pores and micro-cracks from the micrographs. Fig. 6(a) shows the distribution of the pore (crack) area. In total over 49000 cracks and over 200000 pores were identified. The number of cracks shown in Fig. 6(a) was multiplied by 5. There is much more tiny pores (cracks) than large pores (cracks). However, some pixel noise could be misidentified as tiny pores (or cracks). Furthermore, it is reported that the number of tiny pores could be easily altered by changing the greyscale threshold value [2]. Therefore, the number of tiny pores (cracks) shown in Fig. 6(a) could be misleading. Fortunately, these tiny pores are reported to have a limited contribution to porosity [2].

The porosity's distribution on pore (crack) area ($A$) is shown in Fig. 6(b). In this study, the porosity ($S$) is defined as the ratio of sum of each pore's (crack's) area to the whole image area. Because the pore (crack) area varied from 0.001 to 1000 $\mu m^2$, the vertical coordinate of Fig. 6(b) uses d$S$/d(log$A$) instead of d$S$/d$A$. It can be seen that the porosity's distribution has peaks at ~0.1 $\mu m^2$ and ~300 $\mu m^2$ for micro-cracks and pores respectively, far away from the peaks of pore numbers shown in Fig. 6(a). Because the porosity is mainly contributed by relatively large pores (cracks), the porosity's distribution was proved to be not sensitive to the selection of the greyscale threshold value [2]. Although the micro images of one graphite grade may be recorded by different microscopes and analyzed by different researchers, a stable and convincing porosity distribution can be obtained.

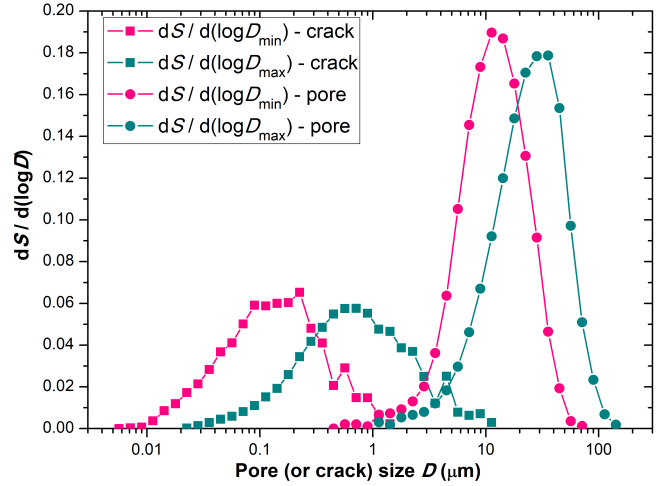

Fig. 7 The porosity's distribution on pore (crack) size.

A ellipse with the same moments of order 1 and 2 as the gas-escape pore or micro-crack was used to define the sizes of the pores or cracks. The lengths of major axis and minor axis of the ellipse are denoted as $D_{max}$ and $D_{min}$ of the pore (crack) respectively. The porosity's distributions on $D_{max}$ and $D_{min}$ are shown in Fig. 7. The peaks are located at ~0.15 μm and ~13 μm for $D_{min}$ while the peaks are located at ~0.7 μm and ~30 μm for $D_{max}$. For micro-cracks, the $D_{max}$ and $D_{min}$ represents the length and width of the crack respectively. For pores with irregular shapes, the $D_{max}$ and $D_{min}$ only represents approximate sizes of the pores. For 3D pore structures reconstructed from X-ray tomography, the diameter of inscribed sphere of one pore was defined as the pore size and used to produce the porosity's distribution on pore size [3]. Using $D_{min}$ defined in this study, the $dS/d(\log D_{min})$ of IG-110 graphite has been demonstrated be to consistent with porosity's distribution derived from X-ray tomography [2]. In other words, the $D_{min}$ is comparable with the diameter of inscribed sphere of graphite pores.

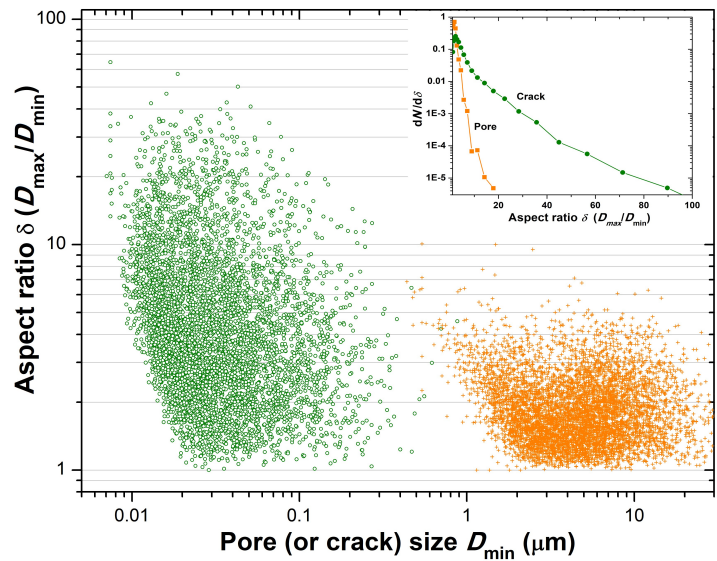

Fig. 8 The aspect ratio ($\delta$) of pore (crack) vs $D_{min}$. The inset shows the distribution of the aspect ratio.

The aspect ratio ($\delta$) of a pore (crack) is defined as $D_{max}/D_{min}$. The aspect ratio vs $D_{min}$ is shown in Fig. 8. In order to explicitly show the distribution of aspect ratio, only one in seven cracks and one in twenty pores were randomly selected and shown in Fig. 8. Most of the pores have aspect ratio smaller than 4. In contrast, the aspect ratio of micro-cracks could be as high as 40. The inset of Fig. 8 shows the distribution of aspect ratio of pores and cracks.

At last, a porosity of 14.2% is obtained in this study for IG-110 gas-escape pores, which is consistent with previous studies [1,2]. A porosity of 3.0% is obtained for IG-110 micro-cracks. The apparent density of the IG-110 sample was measured to be 1.771 g/cm$^3$. Given the ideal graphite density of 2.26 g/cm$^3$, the total porosity is calculated to be 21.6%. There is a gap between the total porosity (21.6%) and the sum (17.2%) of porosity values obtained for pores and micro-cracks, which can be explained as follows. If the graphite sheets in one crystallite are happened to be nearly parallel to the sample's surface, the lenticular micro-cracks will be shown as very shallow dishes on the surface and cannot be identified from the surface image. These missing micro-cracks exposed large cross-sectional area on the sample's surface, and have a considerable contribution to the calculated porosity. In addition, Only large QI particles can be seen at the magnification of 5000×. The tiny cracks in the QI particles shown in Fig. 5(b) cannot be captured at this magnification. Therefore, the true porosity of micro-cracks is actually underestimated by using the image processing method. However, the porosity's distributions shown in Fig. 6 and Fig. 7 are believed to be reliable and could be used as the input data for simulation of graphite behaviors.

## 4. Conclusions

In summary, A three-step polishing method is proposed for graphite materials. The damage induced by each step is identified. Mechanical polishing produced a layer of broken graphite pieces which evidently raised the Raman D mode. These broken graphite pieces were removed by ion milling which also created a disordered surface layer by irradiation in the meantime. A low density carbon material which was reported to fill graphite micro-cracks in previous studies is proved to be this disordered surface layer which in fact covers the micro-cracks. A rapid oxidation process was then adopted to remove this irradiation-induced damage layer and finally exposed true graphite microstructures. This method is recommended to prepare graphite samples for surface characterizations and measurements.

The micro-cracks were observed on rapid-oxidized IG-110 graphite via SEM. By digital image processing, a statistical analysis of IG-110 graphite micro-cracks was conducted. Combining the analysis of micro-cracks and gas-escape pores, the porosity's distribution on crack (pore) size (spanning from 10 nm to 100 μm) is given in this study. it is found that IG-110 graphite's porosity is mainly contributed by cracks with widths of ~0.15 μm and pores with sizes of ~13 μm. The porosity's distribution is valuable for simulation and prediction of graphite performance in reactors.

**Declaration of Competing Interest**

The authors declare that they have no known competing financial interests or personal relationships that could have appeared to influence the work reported in this paper.

**CRediT authorship contribution statement**

**Qing Huang**: Conceptualization, Methodology, Writing - original draft, Investigation, Funding acquisition. **Hui Tang**: Formal analysis, Investigation, Writing - review & editing.

**Acknowledgement**

This work was supported by Youth Innovation Promotion Association of the Chinese Academy of Sciences (grant number 2019262).

**Data availability**

The raw/processed data required to reproduce these findings are available from the authors upon request.